\documentclass[preprint]{aastex63}

\usepackage{aas_macros}
\usepackage{amssymb}
\usepackage{amsmath}
\usepackage{xcolor}

\usepackage{booktabs}
\usepackage{natbib}

\newcommand{\ud}{\mathrm{d}}

\shorttitle{Preprint}

\begin{document}

\title{Metamorphosis of dwarf halo density profile under dark matter decays}

\correspondingauthor{Jianxiong Chen}
\email{jxchen91@link.cuhk.edu.hk}

\author{Jianxiong Chen}
\affiliation{Department of Physics and Institute of Theoretical Physics,\\
The Chinese University of Hong Kong,\\
Shatin, N.T., Hong Kong SAR, China}

\author{M.-C. Chu}
\affiliation{Department of Physics and Institute of Theoretical Physics,\\
The Chinese University of Hong Kong,\\
Shatin, N.T., Hong Kong SAR, China}



\begin{abstract}
We study the density profile of a dwarf halo in the decaying dark matter (DDM) cosmology, using a new algorithm that resolves halo density profiles down to the innermost $700$ pc robustly with high efficiency. Following Schwarzschild's orbit-based method, we have also developed a simplified model to calculate the DDM halo density profiles, which agree remarkably well with those from N-body simulations. Both zoom-in simulations and the simplified model reveal that dark matter decays lead to the flattening of central density and overall reduction of density in dwarf halos, and the underlying physics mechanisms are well illustrated by the simplified model. The slowly-rising scaled rotation curves of DDM dwarf halos agree with the observation of local dwarf galaxies. Our results suggest that the DDM holds great potential for resolving the small-scale problems faced by the cold dark matter (CDM) model.      
\end{abstract}

\keywords{Dark matter (353), N-body simulations (1083), Dwarf galaxies (416), Cosmology (343), Stellar dynamics (1596)}


%
\section{Introduction}
Dark matter (DM) is a crucial ingredient of modern cosmology. It is needed to account for the anisotropy of the cosmic microwave background (CMB)~\citep{wright1992,spergel2003}, large-scale structure formation and evolution~\citep[see, e.g.,][]{springel2006}, dynamics of galaxy clusters and groups~\citep[see, e.g.,][]{zwicky1933}, flattened rotation curves of late-type galaxies~\citep[see, e.g.,][]{bosma1981}, and mismatch between mass centers revealed by gravitational lensing and X-ray emission respectively~\citep[see, e.g.,][]{markevitch2002}. While the DM mass density is about $6$ times higher than that of ordinary matter~\citep{planck2018a}, its nature remains a mystery. Regardless of its composition, the microscopic properties of DM directly affect its cosmic mass distribution. Therefore, through a comparison between theoretical calculation and observation on DM distributions, properties of DM can be constrained. From 1970s, N-body simulation has become an indispensable theoretical tool for dealing with DM's nonlinear clustering. In 1980s, N-body simulations were used to rule out a neutrino-dominated Universe~\citep{white1983} and establish the canonical cold dark matter (CDM) model~\citep{peebles1982a, blumenthal1984, frenk1985, davis1985}. With the improving resolution of N-body simulations, the inner structure of CDM halos were first resolved in 1990s~\citep{dubinski1991, nfw1996}. These high-resolution simulations revealed a central cusp in CDM halos, characterized by a power-law density profile $\rho(r)\propto r^{-\alpha}$ with $\alpha=1$, independent of the halo mass. The cuspy density profile results in a steeply rising rotation curve as $V_{\rm{cir}}\propto r^{1/2}$, which is incompatible with the observed solid-body rotation , $V_{\rm{cir}}\propto r$, of dark matter-dominated dwarf and low surface brightness (LSB) galaxies~\citep{flores1994, moore1994, moore1999}. This tension is referred to as the \emph{core-cusp} problem~\citep{blok2010}. Together with other small-scale problems it challenges the standard CDM model~\citep{bullock2017}.\

Within the CDM framework, the resolution of the \emph{core-cusp} problem relies on baryonic feedbacks. The idea is that the energy released from starburst radiation or supernovae explosions is indirectly transferred to DM particles via frequent fluctuations of potential well driven by repeated and fast gas outflows from the galactic centre~\citep{navarro1996b, pontzen2012}. The extra energy gained by DM particles expands their orbits and converts an initial cuspy density profile into a cored one. The degree of conversion depends on the total energy released from feedbacks measured by stellar mass $M_{\star}$, the total baryonic matter $M_{\rm{b}}$ opposing the expansion, and the energy transfer efficiency. However, the last one is found to be very sensitive to the star formation gas density threshold $n$, a numerical parameter commonly adopted in sub-grid models of galaxy formation~\citep{ben2019, dutton2019}. Cosmological simulations with a large $n$ report core formation in simulated dwarf galaxies~\citep{governato2010, tollet2016,chan2015} while the dark matter cusp remains in simulations using a smaller $n$~\citep{sawala2016}. Till now, a consensus has not been reached about whether baryonic processes can solve the \emph{core-cusp} problem, and it continues to cast doubt on the CDM model.\
 
Apart from the cored profile, many dwarf and LSB galaxies also have similar shape of rotation curves, implying a self-similar dark matter structure~\citep{kravtsov1998, salucci2000}. This conformity is unlikely to be due to chaotic and dramatic baryonic processes~\citep{burkert1995, navarro2019}, but rather a clue to the nature of dark matter beyond CDM. In this work we study a decaying dark matter (DDM) model, which describes two-body decays of DM: 
\begin{equation}\label{eq:ddm_model}
                                           \psi^{\ast} \to \psi + l,
\end{equation} 
where $\psi^{\ast}$ stands for the unstable mother DM particle, $\psi$ is a massive stable daughter DM particle and $l$ is a light and relativistic particle. The DM has therefore multiple components. The dynamics of the two-body decays in this model are fully controlled by two parameters: the decay rate $\Gamma$ , or half-life $\tau^{\ast} = \ln{2}/\Gamma$, of mother particles and $\epsilon$ the energy of $l$ in unit of the mother particle's rest mass. The recoil velocity $V_{k}$ of daughter particles in the center-of-mass frame of their mothers can be expressed as $V_{k} = \epsilon c$, where $c$ is the speed of light.\

There are diverse behaviours of the DDM model depending on the values of $V_{k}$ and $\tau^{\ast}$. For a very small $\tau^{\ast}$ such that almost all mother particles decay into daughter particles before any non-linear structures have formed, the DDM model is similar to the warm dark matter (WDM) model with a free-streaming length determined by the recoil velocity $V_{k}$~\citep{kaplinghat2005, strigari2007}. For a large $V_{k}$ such that the decays convert a fair amount of energy from the matter component to the relativistic species, the expansion history of the whole Universe will be altered~\citep[see, e.g.,][]{vattis2019}. For a broad range of $\tau^{\ast}$ while $V_{k} \lesssim 40$~km~s$^{-1}$, the 1D Lyman-$\alpha$ forest power spectrum predicted by the DDM model is shown to be consistent with observation data~\citep{meiyu2013}, implying that the model behaves like CDM at large scale. Inside a parameter region outlined by $10.0 \lesssim V_{k} \lesssim 40.0$~km~s$^{-1}$ (or equivalently $3.0\times 10^{-5} \lesssim \epsilon \lesssim 1.3\times 10^{-4}$) and $0.1 \lesssim \tau^{\ast} \lesssim 14$ Gyr, decays significantly heat up DM inside dwarf-sized halos, and are relevant for the small-scale problems of CDM~\citep{abdelqader2008, peter2010c, meiyu2014}. In light of this, we revisit the \emph{core-cusp} problem and study the density profiles of dwarf halos in the DDM model using high-resolution cosmological N-body simuations, which is absent from previous studies. 
\

This paper is structured as follows: we first review previous DDM algorithms for N-body simulations and then introduce our new DDM algorithm, test its performance and calibrate numerical parameters for high-resolution zoom-in simulations in Section~\ref{sc:methodology}. An overview of our highest-resolution simulation suite is given in Section~\ref{sc:hres_set}. Section~\ref{sc:semicore} details our mathematical modelling of the evolution of an isolated halo in the DDM cosmology. We present our main results in Section~\ref{sc:res}, followed by more extensive discussions about the DDM model in Section~\ref{sc:discuss}. We summarize in Section~\ref{sc:sum}.

\section{Methodology}\label{sc:methodology}
\subsection{Overview of DDM N-body algorithms}
In the DDM model, light particle $l$ does not take part in the structure formation directly, and is ignored in the following discussion. For the mother particle $\psi^{\ast}$ and daughter particle $\psi$, their Boltzmann equations are
\begin{equation}\label{eq:bzm_mom}
        \frac{\ud f_{\psi^{\ast}}}{\ud t} = -\frac{\ln2}{\tau^{\ast}}f_{\psi^{\ast}},
\end{equation}
and
\begin{equation}\label{eq:bzm_dau}
      \frac{\ud f_{\psi}}{\ud t} = \frac{\ln2}{4\pi \tau^{\ast}V_k^2}\int f_{\psi^{\ast}}(\boldsymbol{r}, \boldsymbol{v'}, t)\delta(|\boldsymbol{v}-\boldsymbol{v'}| - V_{k})\ud^3\boldsymbol{v'},
\end{equation}
respectively, where $f_{\psi^{\ast}}$ and $f_{\psi}$ are the corresponding phase-space mass densities. The DDM matter density evolution can be simulated by using the N-body method once the collision terms in equations~(\ref{eq:bzm_mom}) and~(\ref{eq:bzm_dau}) are properly handled. The first DDM N-body simulation was presented in \citet[][hereafter PMK10]{peter2010}, where the DDM model~(\ref{eq:ddm_model}) was realized on simulation particle basis. In PMK10, each mother simulation particle has a decay probability. Once chosen for decay, the mother simulation particle is flagged to be a daughter simulation particle and receives a random velocity kick at the same time. This Monte-Carlo sampling of decays is carried out at each simulation timestep. Therefore, the global decay rate $\ln2/\tau^{\ast}$ is sampled continually and precisely for the whole system, while the local decay rate fluctuates around the global value with an amplitude depending on the local number density of mother simulation particles. As time goes on, the total number of mother simulation particles drop and the matter density field becomes more and more nonlinear. It can be seen that the decay sampling precision of PMK10 is not uniform in both space and time domains. It is also an intrinsic challenge for the PMK10 algorithm to resolve the central structures of DM halos due to the limited number of mother simulation particles there.\

To achieve a uniform decay sampling in both space and time domains,~\citet[][hereafter CCT15]{dalong2015} proposed a DDM algorithm based on a discretization of the Boltzmann equations~(\ref{eq:bzm_mom}) and~(\ref{eq:bzm_dau}). In CCT15, decays are only sampled at several simulation timings, when each mother simulation particle is split partly to generate a new daughter simulation particle which is kicked randomly at its birth. At other simulation times, all simulation particles are evolved according to the collisionless Boltzmann equation:
\begin{equation}
                      \frac{\ud f_{\psi^{\ast}(\psi)}}{\ud t} = 0.
\end{equation}
The number of mother particles is kept unchanged, and the mass-splitting procedure is the same for all mother simulation particles. Therefore, the decays can be sampled uniformly both in space and time domains. Two numerical parameters are introduced in this algorithm: $f_s$, the number of mass-splittings throughout a simulation, and $N_s$, the number of daughter simulation particles produced per mother simulation particle at each mass-splitting.\

Consider a simulation following the CCT15 algorithm with $f_s$ splittings. Initially it has $N$ mother simulation particles. As each mass-splitting procedure generates $N_sN$ daughter simulation particles, the total number of daughter simulation particles will increase to $f_sN_sN$ by the end of the simulation. For typical values that achieve satisfactory numerical convergence, such as $f_s = 10$ and $N_s = 1$, the final number of daughter simulation particles is larger than that of mother simulation particles by an order of magnitude. Generally the number of simulation particles serves as a measure of the precision of an N-body simulation. However this measure does not apply to the CCT15 algorithm, because the phase space distribution of daughter particles is inferred from $N$ discrete mother simulation particles, not from an underlying continuous distribution function. The sampling resolution of daughter simulation particles is limited by that of the mother simulation particles, which is set in the initial condition. This is also true for the PMK10 algorithm. Hence, increasing the number of daughter simulation particles contributes little to improving the simulation's resolution. On the other hand, the large demand for computing resources of the CCT15 algorithm hinders its practical usage in large cosmological simulations. Therefore improvement of the CCT15 algorithm is needed for our purpose of running high resolution zoom-in simulations.\

\subsection{An improved DDM algorithm}
We follow the framework outlined in CCT15. In our algorithm, the mother simulation particles decay and give birth to daughter simulation particles only at a limited number of decay instances, called breakpoints. The breakpoints are ordered on the time axis in a way that in each phase, the time interval between two adjacent breakpoints, the same number of mother simulation particles decay into daughter simulation particles. When a breakpoint is reached, each mother simulation particle is splitted into a less-massive mother simulation particle and a daughter simulation particle according to the decayed fraction. The newly generated daughter simulation particles are called auxiliary daughters. The mass-splitting procedure increases the total number of simulation particles by $N$. To save memory as well as to speed up the whole simulation, we only track the motion of these $N$ auxiliary daughters for each phase. When the simulation arrives at the next breakpoint, the auxiliary daughters born at the last breakpoint will undergo a random selection process such that only a fraction of them $\eta$ survive and are renamed as permanent daughters. The remaining auxiliary daughter particles are eliminated so that the memory occupied by them is released. All permanent daughters will be traced to the end of the simulation. Therefore there are two states of the daughter simulation particles in our scheme: auxiliary and permanent.\

The auxiliary daughters help to conserve the local matter mass when a decay occurs, a basic conservation observed in both CCT15 and PMK10. After diffusing into the environment, these auxiliary daughters are replaced by permanent daughters, which have a smaller population, hence heavier. The transition from the auxiliary state to permanent state decreases the resolution of daughter particles. Though numerical side-effects can be introduced from the resolution degradation, it is controllable by tuning the survival fraction $\eta$. Suppose there are $N$ mother simulation particles initially and they go through $f_s$ breakpoints. When the simulation is finished, there are $n_fN$ permanent daughters in total, with $n_f = \eta f_s$. Setting $\eta = 1$ brings our algorithm back to CCT15. We implement this algorithm in an individual module named DDMPLUGIN, see Appendix~\ref{app:ddmplugin} for details. It is designed to be compatible with other N-body codes such that the physics associated with dark matter decay is self-contained. This is achieved as follows: during each phase, the whole system is evolved by an N-body CDM code. When a breakpoint is reached, the system's final state is output to the DDMPLUGIN module, which implement the physical effects of dark matter decays. Hence an updated system state is generated. The DDMPLUGIN module then outputs this new state to be the initial condition for next phase's evolution, which is again tracked by the N-body CDM code. Therefore our DDM simulation can be easily implemented in any N-body code, an advantage rooted in the CCT15 algorithm.

\subsection{Cosmological zoom-in simulation}
To study the density profiles of dark matter halos in the DDM cosmology, we run cosmological zoom-in simulations~\citep{onorbe2014} using the DDMPLUGIN module with the N-body code P-GADGET3, a descendant of the public TreePM code GADGET2~\citep{springel2005}. In all simulations, a flat geometry with a cosmological constant is assumed for the background cosmology, where the cosmological parameters are taken from the final results of Planck TT,TE, EE$+$lowE measurements: $\Omega_{m} = 0.3166$, $\Omega_{\Lambda}=0.6834$, $h=0.6727$, $n_s=0.9649$ and $\sigma_8=0.8120$~\citep{planck2018}. Initial conditions are generated by using the code MUSIC~\citep{oliver2011} with the BBKS transfer function~\citep{bbks1986}. The uncertainties induced by the choice of transfer functions are quantified in Appendix \ref{app:cts}. The effect turns out to be negligible. Dark matter halos are identified by the halo finder AHF~\citep{gill2004,knollmann2009}, with Bryan and Norman's fitting formula~\citep{bryan1998} for overdensity $\Delta_c$ calculation. For our adopted cosmology, dark matter halos at redshift $z=0$ are defined by the virial radius $R_{vir}$ within which the mean overdensity is about 103.4 times the critical density $\rho_{crit}$. Based on the position of halo centre, the virial radius, and the best-fit Navarro-Frenk-White (NFW) profile provided by AHF, we collect all particles bound to the target halo and use our own code to measure its radial mass distribution $\bar{\rho}(r)$, the average matter density inside radius $r$. The two-body relaxation time $t_{rel}$ is estimated for certain radii using the method documented in~\citet{binney2008}. Only those radii with $t_{rel}$ larger than the Hubble time $H_{0}^{-1}$ are considered to be reliably resolved~\citep{fukushige2001}.\

\begin{figure}
\centering
\includegraphics[width=0.7\linewidth]{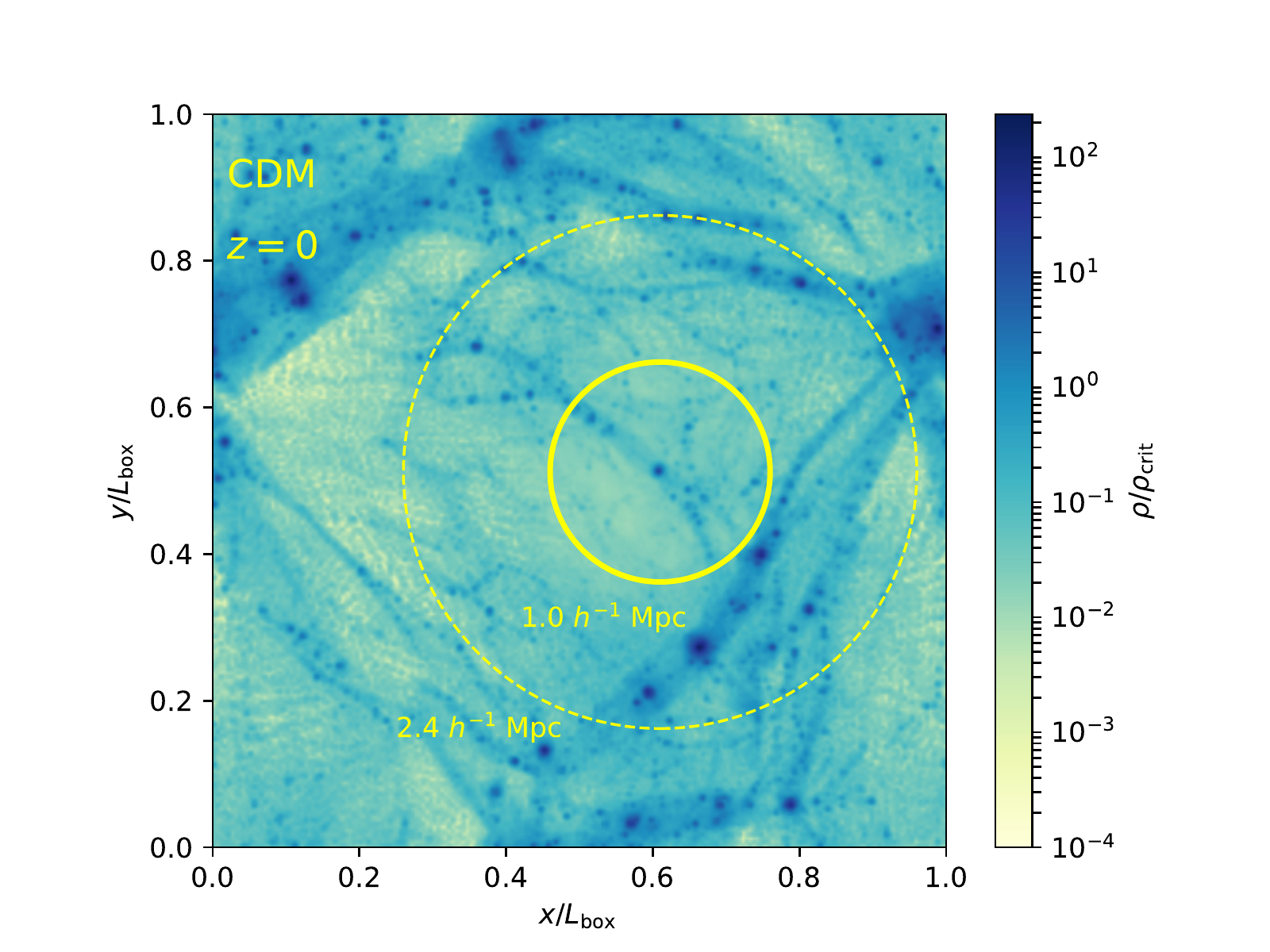}  
\caption{Large-scale environment of the selected dark matter halo, shown in a matter density map of a slice that contains the selected halo at $z=0$. The depth of the slice is $0.1$ times the parent simulation box size. The color encodes the local matter density in units of critical density $\rho_{\mathrm{crit}}$. The two yellow circles are centred at the selected halo, with radii $1.0$~$h^{-1}$ Mpc (solid) and $2.4$~$h^{-1}$ Mpc (dotted). Within this sheet, no structures larger than galaxy-sized halo are found within $1.0$~$h^{-1}$ Mpc range. All cluster-sized dark matter halos are outside $2.4$~$h^{-1}$ Mpc. The data for this plot is taken from a level-12 CDM zoom-in simulation centred at the selected halo.}
\label{fig:lss}
\end{figure}

To select a candidate halo for zoom-in simulations, we run a full-box CDM cosmological simulation starting from redshift $99$. It uses $256^3$ simulation particles inside a periodic cubic box with a width of $6.73$~$h^{-1}$~Mpc, corresponding to a physical length of $10.0$ Mpc at $z=0$. Each CDM simulation particle has a mass of $1.59\times 10^6$ $h^{-1}$ $M_{\odot}$, which is the coarse resolution of the full-box simulation. We select dark matter halos at redshift $0$ by two criteria: small Lagrangian volume and being isolated from larger structures. The selected dark matter halo has a virial mass $M_{\mathrm{vir}} = 5.17\times 10^9$ $h^{-1}$ $M_{\odot}$ and an NFW concentration $c_{\mathrm{vir}} = 21.6$, closely following the theoretically motivated $c-M$ relation of \citet{diemer2018}. Its large-scale environment is illustrated in Figure~\ref{fig:lss}, showing that the selected halo is far away from surrounding larger structures. \

For zoom-in simulations centred at the selected halo, the initial density fluctuations inside its Lagrangian volume are sampled by high-resolution particles while its large-scale environment is represented by coarse resolution particles. A buffer volume is created between the above two regions to avoid large resolution gradient. We label the resolution level of a zoom-in simulation by an integer $l$. A level-$l$ zoom-in simulation has a mass resolution equivalent to that of a full-box N-body simulation using $(2^l)^3$ CDM particles in its initial condition. We adopt the empirical formula recommended by \citet{power2003} to calculate the gravitational softening lengths for high-resolution CDM zoom-in simulation particles. As for DDM zoom-in simulations, the decays of mother simulation particles are only switched on inside the high-resolution volume. It is reasonable since we do not consider decay parameters that result in significant deviations from the CDM large-scale matter distribution. Gravitational softening lengths for high-resolution DDM particles are set to be the same as those of CDM zoom-in particles which have the same resolution level. This reflects the fact that the resolution of a DDM simulation is constrained by its initial condition. We also require that zoom-in halos are free of contamination by lower resolution particles within their virial radii for a clean analysis. In the following subsection, we test and calibrate $f_s$ and $n_f$ of our DDM algorithm for dark matter halo density profile study.

\subsection{Numerical parameters calibration}\label{sc:npcalib}
\begin{figure}[t]
\centering
\includegraphics[width=0.45\linewidth]{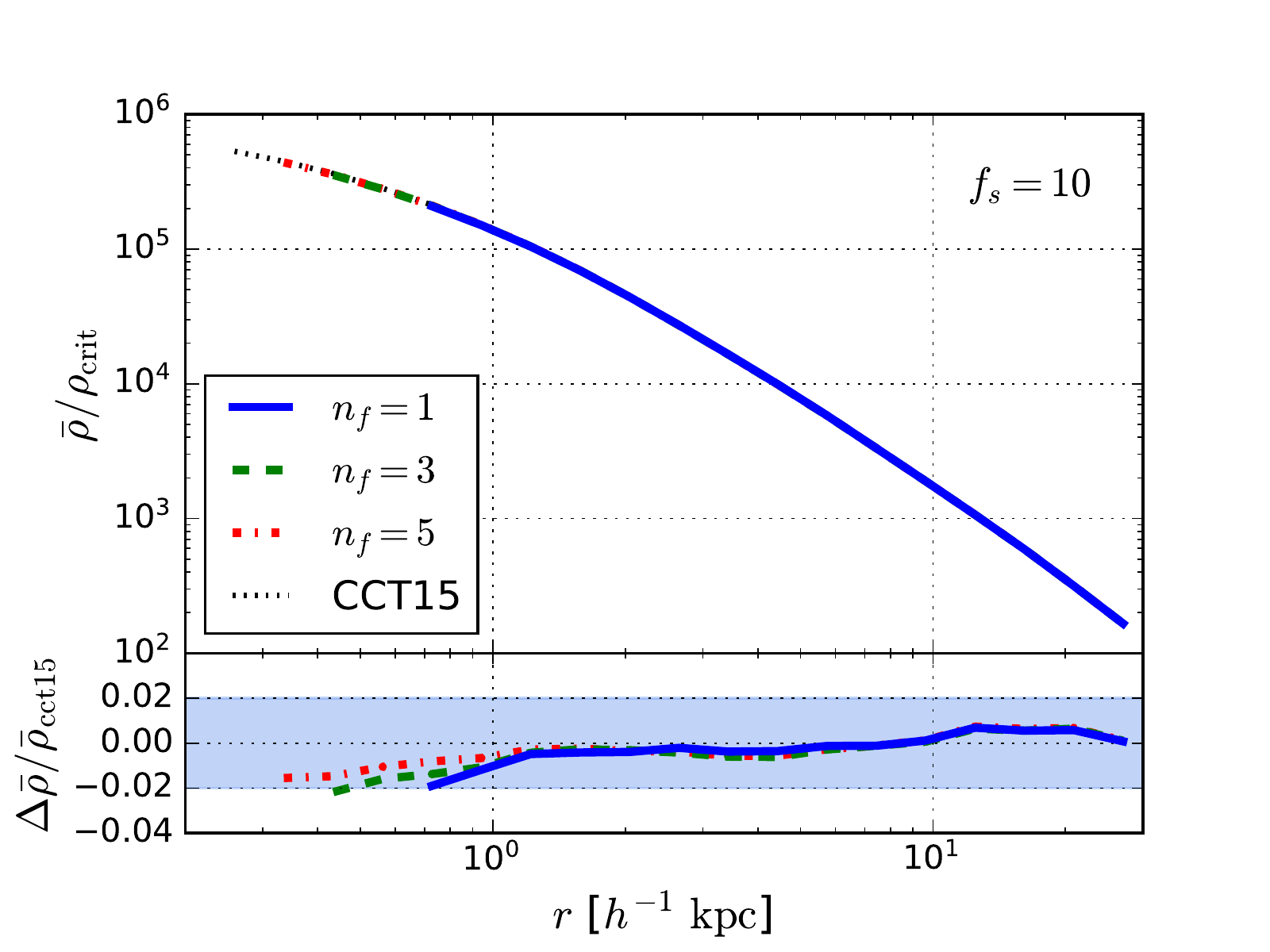}  
\includegraphics[width=0.45\linewidth]{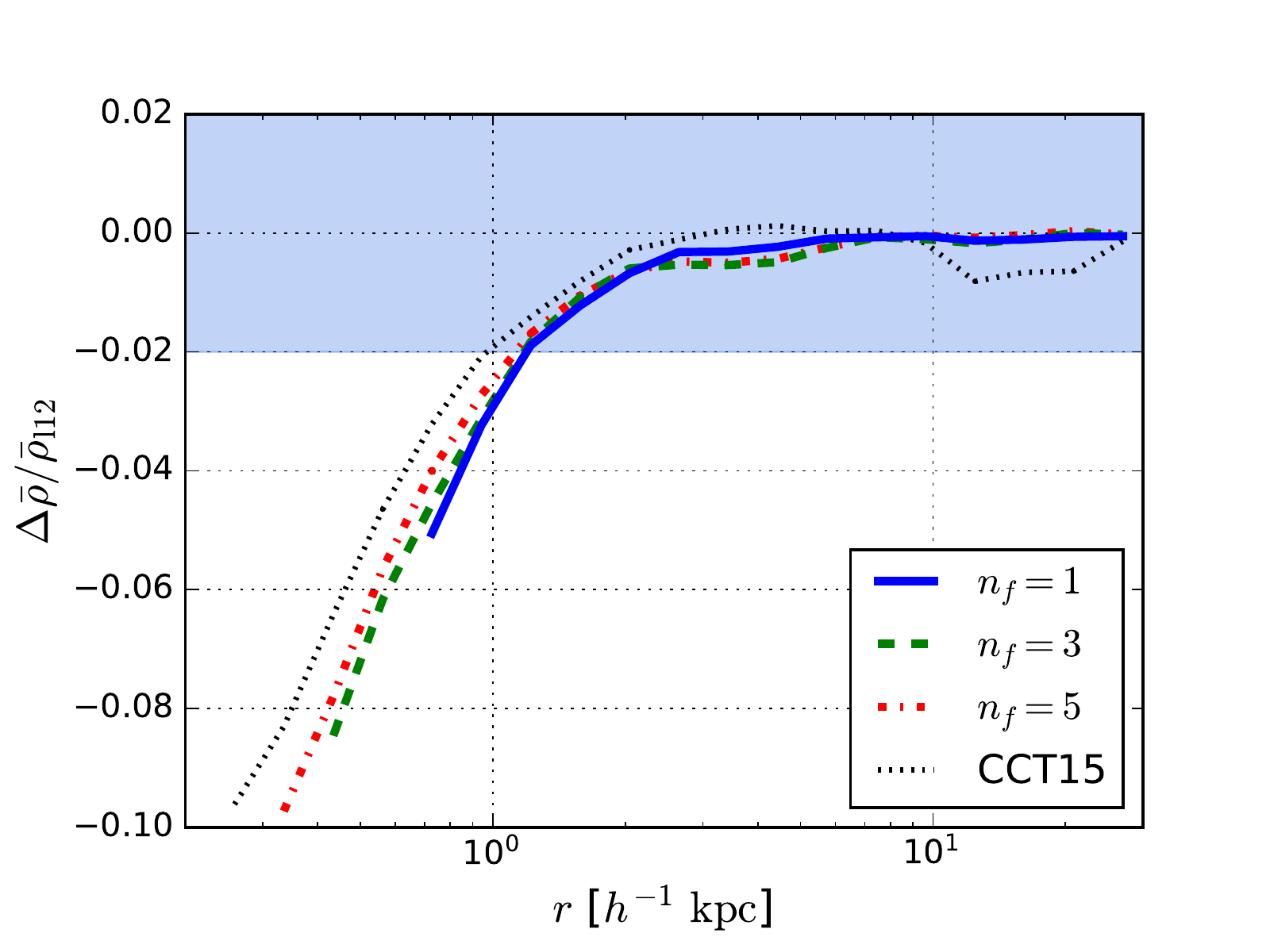}  
\caption{Effects of the numerical parameter $n_f$ on the halo average density profile $\bar{\rho}(r)$. All profiles are measured at redshift $0$ and all DDM zoom-in simulations use the same decay parameters: $V_k = 20.0$~km~s$^{-1}$ and $\tau^{\ast} = 3.0$ Gyr. Results are plotted for $r \geqslant r_{\mathrm{rel}}$ where the two-body relaxation time $t_{\mathrm{rel}}$ reaches the Hubble time $H_0^{-1}$. For both the left and right panels, the blue solid, green dashed, red dot-dashed and black dotted lines plot the results of the simulation using $n_f = 1$, $3$, $5$ and $10$, the last of which is equivalent to that of the CCT15 algorithm, respectively. All these $4$ zoom-in simulations are carried out at level $11$. The left panel shows the density profiles of all level-$11$ runs and their relative differences to the profile from the CCT15 simulation. For $n_f=1$, we also run a level-$12$ simulation. The right panel shows the relative differences between all level-$11$ density profiles and the level-$12$ density profile.}
\label{fig:nf-test}
\end{figure}

The parameter $f_s$ controls the decay sampling frequency, and a larger value of it leads to a finer sampling of the decay history. In CCT15, the total number of daughter particles is linearly proportional to $f_s$. Hence it cannot be arbitrarily large, otherwise the computational work load will be huge. However this constraint on $f_s$ is released in our algorithm as the total number of daughter particles is solely determined by $n_f$. For DDM zoom-in simulations used in this study, we set $f_s = 10$ by default. It proves to be large enough for acceptable numerical convergences (see Appendix~\ref{app:fs_test} for details) .\

In our algorithm, $n_f$ controls the total number of simulation particles. To test its effects on the halo average density profile, we tried three values: $n_f=1, 3,$ and $5$. A simulation using the CCT15 algorithm (equivalent to $n_f = f_s$) is also carried out for comparison. All simulations are run in level-$11$ resolution using the same decay parameters $V_k = 20.0$~km~s$^{-1}$ and $\tau^{\ast} = 3.0$ Gyr. The density profiles at $z=0$ are measured down to the inner-most resolved radii and are shown in the left panel of Figure~\ref{fig:nf-test}. The density profiles with different values of $n_f$ all converge to CCT15's density profile within $2\%$ at all resolved radii. It implies that the radial mass distribution $\bar{\rho}(r)$ is insensitive to the value of $n_f$ when $n_f \geqslant 1$. We further run a level-$12$ simulation using the same decay parameters for $n_f = 1$. In the right panel of Figure~\ref{fig:nf-test}, we compare all level-$11$ profiles with the level-$12$ one. For $r \lesssim 1.0$ $h^{-1}$ kpc, all level-$11$ profiles are systematically lower than the level-$12$ profile. The differences continue to grow when $r$ approaches to the halo centre. A large $n_f$ does help to narrow the differences down, however the cost is also huge: the gain in precision is only about $3\%$ by increasing $n_f$ from $1$ to $10$ (the CCT15 algorithm). As a larger value of $n_f$ brings forth a larger number of simulation particles, the two-body relaxation converged radius $r_{\mathrm{rel}}$ decreases as the value of $n_f$ increases. However, for $n_f > 1$, $r_{\mathrm{rel}}$ cannot be taken as the inner-most resolved radii anymore, because the deviation from the level-$12$ profile at $r_{\mathrm{rel}}$ also grows with the $n_f$ and reaches $10\%$ level for $n_f \geqslant 5$. Hence for the level-$11$ runs, increasing the value of $n_f$ does not decrease the inner-most resolved radius reliably. On the contrary, the inner-most resolved limit set by the $r_{\mathrm{rel}}$ of the $n_f=1$ run is worthy: the density profile of the level-$11$ resolution converges to that of the level-$12$ resolution within $5\%$ for all radii larger than $0.8$ $h^{-1}$ kpc. Then we can safely use $n_f=1$ without worrying about possible degradation of resolution and take $r_{\mathrm{rel}}$ as the inner-most resolved radius. For our highest resolution zoom-in simulations, the level-$12$ runs, we use $n_f = 1$ and $f_s=10$. The results are presented in the next section.


\section{High-resolution zoom-in simulations}\label{sc:hres_set}

\begin{table}[t]
\caption{Halo properties of our simulation suite, including the name of each level-$12$ zoom-in simulation (column $1$), recoil velocity of daughter particles (column $2$), decay half-life (column $3$), virial mass (column $4$), virial radius (column $5$), inner-most resolved radius (column $6$), characteristic scale (column $7$) and the corresponding characteristic velocity (column $8$), concentration obtained from an NFW fitting (column $9$), and total particle number inside the virial radius (column $10$). See Section~\ref{sc:res} for details.}
\label{tab:halo_property}
\def\arraystretch{1.2}
\hspace*{-1.6cm}
\begin{tabular}{lccccccccr}
\midrule
\midrule
 \multicolumn{1}{l}{  Name  }&
    \multicolumn{1}{c}{  $V_k$  }&
    \multicolumn{1}{c}{  $\tau^{\ast}$ }&
    \multicolumn{1}{c}{  $M_{\rm{vir}}$  }&
    \multicolumn{1}{c}{  $R_{\rm{vir}}$  }&
    \multicolumn{1}{c}{  $r_{\rm{rel}}$  }&
    \multicolumn{1}{c}{  $R_{0.3}$}&
    \multicolumn{1}{c}{  $V_{0.3}$}&
    \multicolumn{1}{c}{  $c_{\rm{nfw}}$  }&
    \multicolumn{1}{c}{  $N_{p}$ }\\
    
    \multicolumn{1}{c}{  }&
    \multicolumn{1}{c}{  \small{(km~s$^{-1}$)}  }&
    \multicolumn{1}{c}{  \small{$(\rm{Gyr})$} }&
    \multicolumn{1}{c}{  \small{$(10^9$ $h^{-1}$ $M_{\odot})$} }&
    \multicolumn{1}{c}{  \small{($h^{-1}$ kpc)} }&
    \multicolumn{1}{c}{  \small{($h^{-1}$ kpc)} }&
    \multicolumn{1}{c}{  \small{($h^{-1}$ kpc)} }&
    \multicolumn{1}{c}{  \small{(km~s$^{-1}$) }}&
    \multicolumn{1}{c}{  }&
    \multicolumn{1}{c}{  $(10^{7})$  }\\
   
\midrule
  CDM       & ...  & ...     & 5.17   & 35.0  & 0.254    & 0.407    & 27.7   & 21.6  & 1.33    \\
  V20T3     & 20.0  & 3.00    & 4.22   & 32.7  & 0.291    & 0.904    & 26.0   & 15.0  & 2.35    \\
  V20T7     & 20.0  & 6.93    & 4.41   & 33.2  & 0.224    & 0.805    & 27.4   & 17.3  & 2.36    \\
  V20T14    & 20.0  & 14.0    & 4.70   & 33.9  & 0.200    & 0.684    & 28.7   & 19.3  & 2.41    \\
  V30T3     & 30.0  & 3.00    & 2.89   & 28.8  & 0.366    & 1.72     & 20.1   & 6.99  & 1.93    \\
  V30T7     & 30.0  & 6.93    & 3.32   & 30.2  & 0.254    & 1.23     & 23.3   & 11.4  & 1.91    \\
  V30T14    & 30.0  & 14.0    & 4.05   & 32.3  & 0.215    & 0.810    & 26.4   & 16.8  & 2.08    \\
  V40T3     & 40.0  & 3.00    & 0.350  & 14.3  & 0.470    & 3.58     & 10.0   & 4.64  & 0.377   \\
  V40T7     & 40.0  & 6.93    & 1.79   & 24.6  & 0.278    & 1.69     & 18.1   & 9.93  & 1.19    \\
  V40T14    & 40.0  & 14.0    & 3.26   & 30.0  & 0.234    & 1.00     & 23.7   & 13.5  & 1.67    \\
\midrule
\midrule
\end{tabular}
\end{table}

Our level-$12$ simulation suite comprises 10 cosmological zoom-in simulations: 1 CDM and 9 DDM realizations. All 10 simulations use the same initial condition at redshift $99$. The high-resolution volume in the initial condition is a cuboid with three edge-lengths being $0.09$, $0.09$ and $0.15$~$h^{-1}$~Mpc, centering at its parent simulation box. The mass of each high-resolution particle is $3.89\times 10^2$~$h^{-1}$~$M_{\odot}$. The gravitational softening length of high-resolution CDM particles is set to be $33.7$~$h^{-1}$~pc, frozen to a physical length about $50$ pc after $z=10$. All DDM realizations follow the same softening length assignment scheme such that the force resolution of our simulation suite is uniform. All our zoom-in halos are free of contamination by low-resolution particles within their virial radii. The $9$ DDM realizations differ from each other in the combination of $V_k$ and $\tau^{\ast}$. We have used $3$ values of $\tau^{\ast}$: $3.0$, $6.93$ and $14.0$ Gyr, with corresponding decayed fractions $0.959$, $0.748$ and $0.495$, respectively. For each $\tau^{\ast}$, we use $3$ different values of $V_k$: $20.0$, $30.0$ and $40.0$~km~s$^{-1}$. These $9$ realizations constitute a rough sampling of the interesting region in the $\tau^{\ast}-V_k$ parameter space. Halo expansion has been observed in the 9 DDM zoom-in halos as they generally have smaller virial masses but lower concentrations compared to the corresponding CDM halo. The virial masses, virial radii and other global properties of all zoom-in halos are summarized in Table~\ref{tab:halo_property}. We study the physics accounting for the DDM halo expansion in next section.


\section{A simplified semi-analytic model of DDM density profiles}\label{sc:semicore}
The halo expansion in the DDM model is driven by two primary physical processes. The first one is the decay itself. On average, the kinetic energy of newly born daughter particles are greater than those of their mothers. This extra kinetic energy drives the orbits of daughters outwards, hence expanding the whole halo. We call this the Step-$1$ expansion. A consequence of this expansion is the weakening of the gravitational potential, triggering the Step-$2$ expansion: the bulk particles' orbits expand outwards to rebalance the weakened gravity with the inertial force seen in the orbits' rotating frames.
\citet{cen2001} considers a special case of the two-step expansion: $V_k \gg v_e$ and $\tau > t_{\mathrm{dyn}}$, where $v_e$ and $t_{\mathrm{dyn}}$ are the halo's typical escape velocity and dynamical time, respectively. The large $V_k$ unbinds all daughter particles during the Step-$1$ expansion while the slow decay simplifies the Step-$2$ evolution to an adiabatic expansion. Starting from an NFW density profile, the resulting density profile turns out to remain an NFW shape, but with a smaller concentration and a lower normalization density (see also the relavant discussion in \citet{peter2010b}). \citet{sanchez-salcedo2003} considers the situation where most of daughter particles are bound to the halo ($V_k < v_e$) and the halo expands adiabatically ($\tau > t_{\mathrm{dyn}}$). Through simple semi-analytic calculations, he argued that the cored profile is a natural result of the two-step expansion. In this section, we present a general formalism to implement the two-step expansion.

Our model assumes that a dark matter halo forms at a high redshift when only a small fraction of mother particles have decayed. Initially the halo is CDM-like with an NFW density profile. Then decays proceed and the density profile evolves. The decay of a mother particle is a random process and does not have a preferential direction. On average, each newborn daughter particle acquires an additional amount of kinetic energy from the mass deficit of its mother particle:
\begin{equation}\label{eqn:mother_daughter_ek}
            \langle E_{k,\mathrm{dau}}\rangle = E_{k, \mathrm{mom}} + \frac{1}{2}m_{\rm{dau}}V_k^2,
\end{equation} 
where $E_{k, \mathrm{mom}}$ is the kinetic energy of the decayed mother particle and $\langle E_{k,\mathrm{dau}}\rangle$ is the expected kinetic energy of its daughter particle with mass $m_{\rm{dau}}$. Similarly, the mean angular momentum of a daughter particle $\langle\boldsymbol{L_{\rm{dau}}}\rangle$ is the same as that of its mother, $\boldsymbol{L_{\rm{mom}}}$,
\begin{equation}\label{eqn:mother_daughter_am}
                      \langle\boldsymbol{L_{\rm{dau}}}\rangle = \boldsymbol{L_{\rm{mom}}}.
\end{equation}
Based on equation~(\ref{eqn:mother_daughter_ek}) and equation~(\ref{eqn:mother_daughter_am}), we build our simplified model using Schwarzschild's orbit-based method~\citep{schwarzschild1979} (see~\citet{julio2010} for a short overview), which represents a collisionless system by a large library of particle orbits. Physical quantities, such as mass distribution, are then derived through constructing superpositions of these orbits.\

Given a gravitational potential $\Phi(r)$, a particle's orbit is a function of its total energy $E$ and angular momentum $\boldsymbol{L}$. The joint distribution of $E$ and $\boldsymbol{L}$ gives the whole library of particle orbits. To make the model as simple as possible, we assume all mother particles take circular orbits around the halo center, with random orientations of orbital planes. The daughter particles take up rosette orbits. \

Now we consider the enclosed mass profile $M(R)$ as a superposition of orbits. Mathematically, all orbits in the library form a set. We name it as \emph{Olib}. For each element $x$ in \emph{Olib}, a weighting factor $g_x(R)$ is assigned such that $M(R)$ is the summation of all particle masses weighted by $g_{x}(R)$ over the set \emph{Olib}:
\begin{equation}\label{eq:enclosed_mass_as_g_summation}
                             M(R) = \sum_{x\in \emph{Olib}} m_xg_x(R),
\end{equation}
where $m_x$ is the mass of the particle moving in the orbit $x$. The weighting factor $g_x(R)$ can be constructed as the fraction of time the particle spends inside the sphere $r=R$:
\begin{equation}\label{eq:g_def}
                                   g_x(R) = \Delta t_x(R)/T_x,
\end{equation}
where $\Delta t_x(R)$ is the duration that a particle travels inside the sphere $r=R$ within an orbital period $T_x$. The weighting factors for circular orbits $g_{\rm{cir}}(R, r)$ are step functions since a circle with radius $r$ is either totally inside or totally outside the sphere $R$:
\begin{equation}
                                        g_{\rm{cir}}(R, r) =
                                             \begin{cases}
                                             0 & \text{if } r > R,\\
                                             1 & \text{if } r \leqslant R.
                                             \end{cases}
\end{equation}

Similarly for a rosette orbit, if its perihelion $r_{\rm{min}}$ (aphelion $r_{\rm{max}}$) is larger (smaller) than $R$, then its weighting factor $g_{\rm{ros}}(R, r_{\rm{min}}, r_{\rm{max}})$ is 0 (1). If $r_{\rm{min}} < R < r_{\rm{max}}$, the weighting factor is calculated as follows:
\begin{equation}
           g_{\rm{ros}}(R, r_{\rm{min}}, r_{\rm{max}}) = \frac{\int_{r_{\rm{min}}}^{R}\frac{\mathrm{d}r}{\sqrt{E-V_{\rm{eff}}(r)}}}{\int_{r_{\rm{min}}}^{r_{\rm{max}}}\frac{\mathrm{d}r}{\sqrt{E-V_{\rm{eff}}(r)}}},
\end{equation}  
where $V_{\rm{eff}}(r)$ is the effective potential.

Consider the Step-$1$ expansion during a small time interval $\Delta t$ such that the gravitational potential $\Phi(r)$ remains static. Consider a pair of mother and daughter particles: a circular orbit with radius $r_0$ before decay and the corresponding rosette orbit after decay. According to equation~(\ref{eqn:mother_daughter_am}), they share the same specific angular momentum $l_0$, thus the same effective potential:
\begin{equation}
                        V_{\rm{eff}} = \frac{l_{0}^2}{2r^2} + \Phi(r).
\end{equation}
The value of $l_0$ can be obtained by considering the circular motion at radius $r_0$, where $V_{\rm{eff}}$ reaches its minimum. As for the rosette orbit, its perihelion and aphelion are the two roots of the following equation from equation~(\ref{eqn:mother_daughter_ek}):
\begin{equation} \label{eq:rm_roots}
               V_{\rm{eff}}(r) - V_{\rm{eff}}(r_0) - \frac{1}{2}V_{k}^2 = 0,
\end{equation} 
They both depend on the radius $r_0$ of its mother particle's orbit. If equation~(\ref{eq:rm_roots}) has only one root, daughter particle is considered to be unbound and makes no contribution to the mass profile $M$($R$).\

After $\Delta t$, the mass increment of daughter particles inside the sphere $r = R$ can be calculated by considering the contributions from the daughter particles which are born during this time interval:
\begin{equation}
  \Delta M_{\rm{dau}}(R,\Delta t, t_i) = \Delta f\int_0^{\infty}4\pi r^2\rho_{\rm{mom}}(r,t_i)g_{\rm{ros}}\left[R, r_{\rm{min}}(r, V_k), r_{\rm{max}}(r, V_k), t_i\right]\rm{d}r,
\end{equation}
where $\Delta f$ is the fraction of mother particles decayed during $\Delta t$:
\begin{equation}
                               \Delta f = \ln(2)\Delta t/\tau^{\ast}.
\end{equation}
$\rho_{\rm{mom}}(r, t_i)$ is the density profile of mother particles at time $t_i$. The mass profile of mother particles declines uniformly by a factor of $\Delta f$:
\begin{equation}
                 \Delta M_{\rm{mom}}(R,\Delta t, t_i) = -\Delta f M_{\rm{mom}}(R, t_i).
\end{equation}
Then after the Step-$1$ expansion, the total mass profile $M(R)$ changes by an amount of
\begin{equation}
            \Delta M(R, \Delta t, t_i) = \Delta M_{\rm{dau}}(R,\Delta t, t_i)-\Delta f M_{\rm{mom}}(R, t_i).
\end{equation}
We proceed to consider the Step-$2$ expansion in the time interval $(t_i, t_i+\Delta t)$. For the sphere $r = R$ at $t_i$, the adiabatic expansion reads:
\begin{equation}\label{eq:adiabatic}
                  RM(R, t_i) = R_f(t_{i+1})[M(R, t_i)+\Delta M(R, \Delta t, t_i)],
\end{equation}
from which $R_f(t_{i+1})$, the radius at $t_{i+1} = t_i + \Delta t$, can be derived. Initially the total mass inside $R$ is $M(R, t_i)$. After $\Delta t$, the total mass inside $R_f(t_{i+1})$ is $M(R, t_i)+\Delta M(R, \Delta t, t_i)$. Therefore the mass profile $M$($R,t$) evolves a bit.\

Through the two-step expansion we have evolved the whole system from $t_i$ to $t_{i+1}$. We loop the procedures listed above and evolve the whole system from the initial time $t_0$ to the final time $t_f$.

\section{Results}\label{sc:res}
\begin{figure}
\centering
\includegraphics[width=0.7\linewidth]{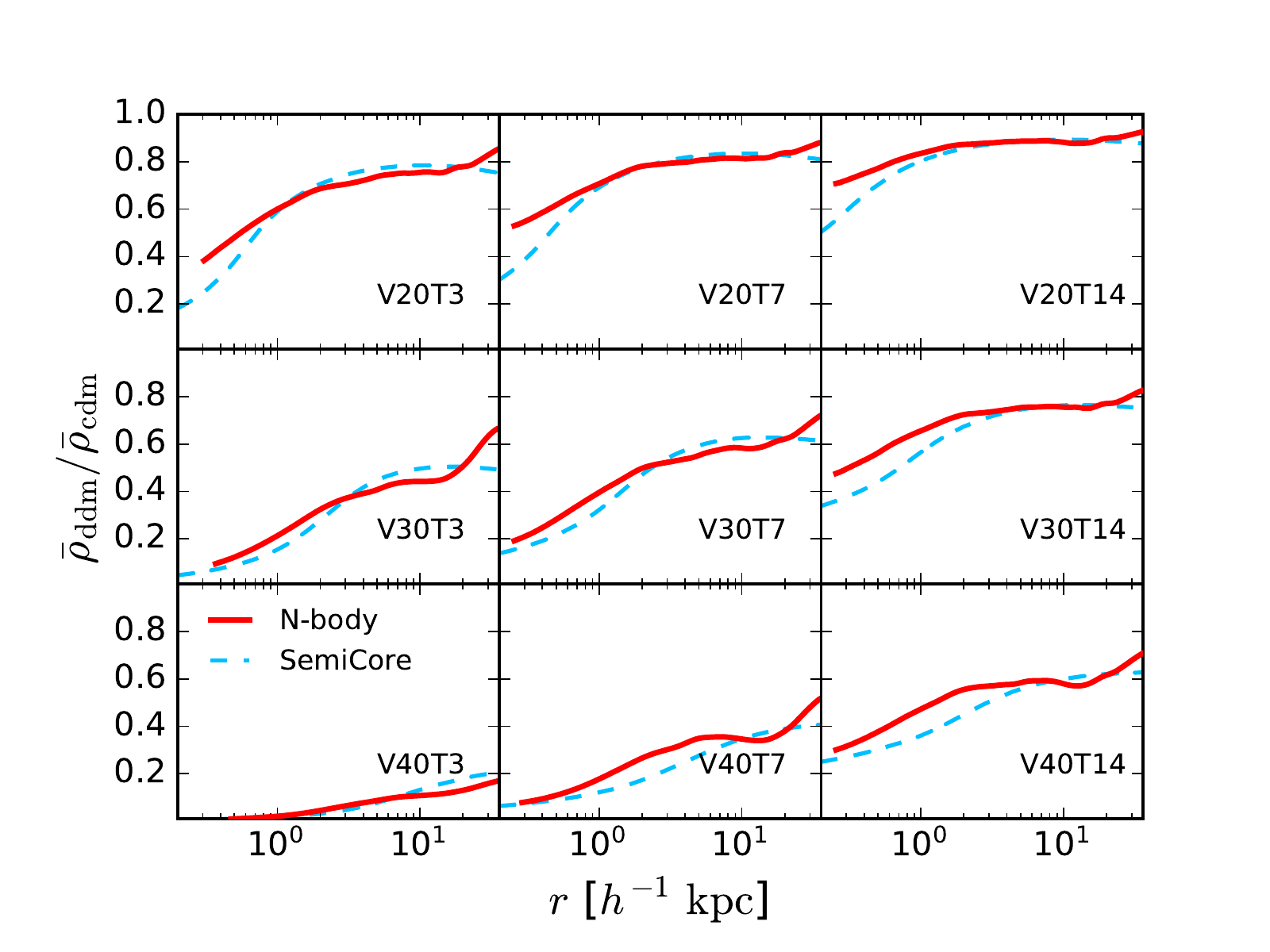}  
\caption{Comparisons between the level-$12$ N-body simulation and the \emph{SemiCore} calculation of the average density ratio $\bar{\rho}_{\rm{ddm}}(r)/\bar{\rho}_{\rm{cdm}}(r)$ between a DDM-CDM halo pair. In each subpanel, the solid red line shows the result of N-body simulations down to the innermost resolved radius $r_{\rm{rel}}$. The light-blue dashed line is the prediction of the \emph{SemiCore} model. The decay parameters used in each subpanel are implied by the DDM halo name (see Table~\ref{tab:halo_property} for details). For all N-body simulation curves, the statistical uncertainties from Poisson noises can be safely neglected for $r\geqslant r_{\rm{rel}}$. The systematic uncertainty of our DDM algorithm introduces a $\sim 6\%$ uncertainty for $\bar{\rho}_{\rm{ddm}}/\bar{\rho}_{\rm{cdm}}$ at $r_{\mathrm{rel}}$. The systematic uncertainties decrease as $r$ increases and become unimportant, see Section~\ref{sc:npcalib} for details.}
\label{fig:averho_ratio}
\end{figure}

The simplified semi-analytic model presented in Section~\ref{sc:semicore} is implemented with a well-tested numerical code called \emph{SemiCore}. Starting from the best-fit NFW profile of the level-$12$ CDM halo ($M_{\rm{vir}} = 5.17\times 10^9$~$h^{-1}$~$M_{\odot}$ and $c_{\mathrm{vir}} = 21.6$), we run \emph{SemiCore} with the combinations of $V_{k}$ and $\tau^{\ast}$ listed in Table~\ref{tab:halo_property}, with the initial and final times being the same as those of cosmological N-body simulations. In Figure~\ref{fig:averho_ratio}, we show the present ($z=0$) average density profile ratios, $\bar{\rho}_{\rm{ddm}}(r)/\bar{\rho}_{\rm{cdm}}(r)$, between the DDM-CDM halo pairs which evolve from the same primeval local overdensity field. We compare the results of the \emph{SemiCore} model with those from the level-$12$ N-body simulations. For the \emph{SemiCore} model, $\bar{\rho}_{\rm{cdm}}(r)$ is calculated from the input NFW profile.\

The simulation data reveals that dark matter decays reduce the mass profile throughout the dwarf halos. The global reduction amplitude increases as $V_{k}$ increases or $\tau^{\ast}$ decreases. For a given pair of decay parameters, the difference between DDM and CDM halos become more pronounced as $r$ approaches to the halo center. These trends are well reproduced by the \emph{SemiCore} model. Furthermore, the average density ratios $\bar{\rho}_{\rm{ddm}}/\bar{\rho}_{\rm{cdm}}$ predicted by \emph{SemiCore} agree with those from sophisticated N-body simulations to better than $40\%$, for all resolved radii and for all combinations of $V_{k}$ and $\tau^{\ast}$ considered in this study. The triumph of \emph{SemiCore} confirms the two-step expansion scenario for explaining the halo expansion induced by dark matter decays.\

Figure~\ref{fig:averho_ratio} also shows that \emph{SemiCore} systematically overpredicts the mass reduction in the inner region ($r \lesssim 0.1R_{\rm{vir}}$). It may be related to the simplified assumption of circular orbits for all mother particles. The same effect was observed in modelling adiabatic contraction of dark matter by circular orbits, which leads to an enhancement of the central density relative to the results of high resolution simulations~\citep{gnedin2004}. In the outer region ($r \gtrsim 0.8R_{\rm{vir}}$), N-body simulations show that the ratios $\bar{\rho}_{\rm{ddm}}/\bar{\rho}_{\rm{cdm}}$ continue to grow while the \emph{SemiCore} model predicts a flattened or slightly declining shape. Notice that the decay of a mother particle in a rosette orbit generally produces daughter particles that reach larger $r$, compared to those from a circular mother orbit. Also, there is mass accretion from the environment in N-body simulations, which is absent in the \emph{SemiCore} model. Both factors can be responsible for the discrepancy seen in the outer halo region.\

\begin{figure}
\centering
\includegraphics[width=0.7\linewidth]{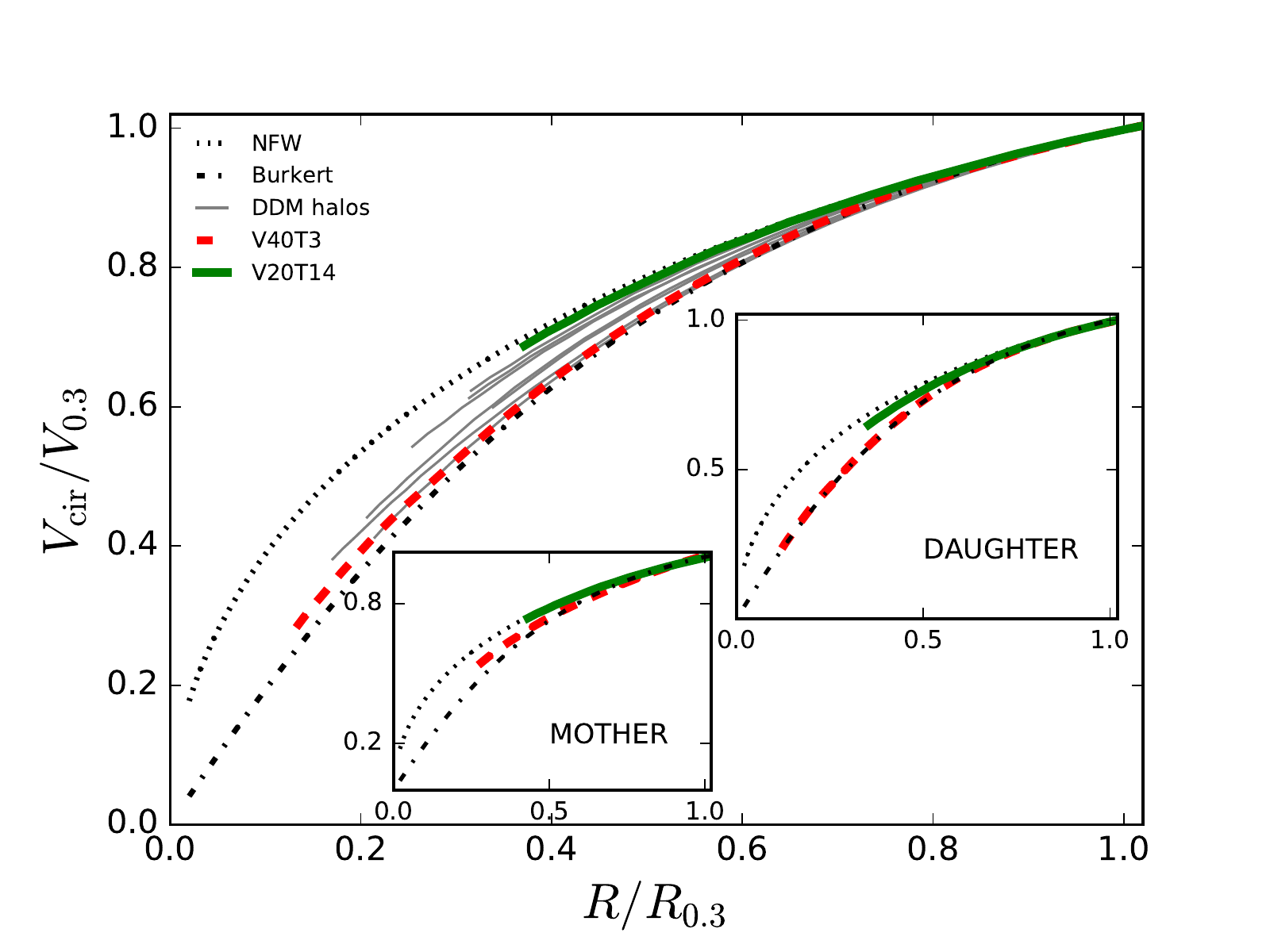}  
\caption{Scaled inner rotation curves of the $9$ level-$12$ DDM zoom-in dwarf halos. The rotation velocities $V_{\rm{cir}}(r)$ are scaled by a characteristic velocity $V_{0.3}$ while the radii are scaled by $R_{0.3}$, where $V_{\rm{cir}} = V_{0.3}$ and $\ud\ln{V_{\rm{cir}}}/\ud\ln{r} = 0.3$. Simulation data are only shown for $r \geqslant r_{\rm{rel}}$. The parent panel shows the results for the total matter field while the two subpanels show the shapes for mother and daughter particles, respectively. The results of $\rm{V}20\rm{T}14$ and $\rm{V}40\rm{T}3$ are shown in solid green lines and dashed red lines, respectively. The results of other simulated halos are only shown in the parent panel using gray lines. Two theoretical curves are also plot for comparison, based on the NFW profile (black dotted line) and Burkert profile (black dot-dashed line).}
\label{fig:sRC_ensemble}
\end{figure}

In Figure~\ref{fig:averho_ratio}, the average density ratios from N-body simulations display a common shape: rising inner and outer regions connected by an extended plateau. A positive slope of the $\bar{\rho}_{\rm{ddm}}/\bar{\rho}_{\rm{cdm}}$ curve implies the flattening of DDM density profile compared to that of its CDM countpart. We calculate the rotation curve $V_{\rm{cir}}(r)$ scaled by $V_{0.3}$, the circular velocity at radius $R_{0.3}$ where $\ud\ln{V_{\rm{cir}}}/\ud\ln{r} = 0.3$~\citep{hayashi2006}. In Figure~\ref{fig:sRC_ensemble}, we plot the scaled rotation curves for all level-$12$ DDM halos, and their values of $R_{0.3}$ and $V_{0.3}$ are listed in Table~\ref{tab:halo_property}. The scaled rotation curves based on the cuspy NFW profile and cored Burkert profile~\citep{burkert1995} are also plotted together for comparison. The DDM curves spread between the two theoretical curves in two groups. One is made up of $4$ DDM halos: $\rm{V}20\rm{T}14$, $\rm{V}30\rm{T}14$, $\rm{V}40\rm{T}14$ and $\rm{V}20\rm{T}7$. Their scaled rotation curves are closer to the NFW curve than the Burkert one. The remaining $5$ halos form the second group which features a significant deviation from the NFW curve and is much closer to the Burkert curve. For $\rm{V}20\rm{T}14$, a member of the pro-NFW group, and $\rm{V}40\rm{T}3$, a member of the pro-Burkert group, we further plot their scaled rotation curves of mother and daughter components in Figure~\ref{fig:sRC_ensemble}. Surprisingly, the scaled daughter (mother) rotation curve of $\rm{V}40\rm{T}3$ ($\rm{V}20\rm{T}14$) follows the Burkert (NFW) shape. Since $\rm{V}40\rm{T}3$ has the strongest decay effect while $\rm{V}20\rm{T}14$ has the weakest, the simulation results show that the halo expansion due to dark matter decays can flatten the halo density profile and transform it from a cuspy shape to the cored shape, depending on the combination of $V_{k}$ and $\tau^{\ast}$ for a given dwarf halo mass.

\section{Discussion}\label{sc:discuss}
\begin{figure}[t]
\centering
\includegraphics[width=0.48\linewidth]{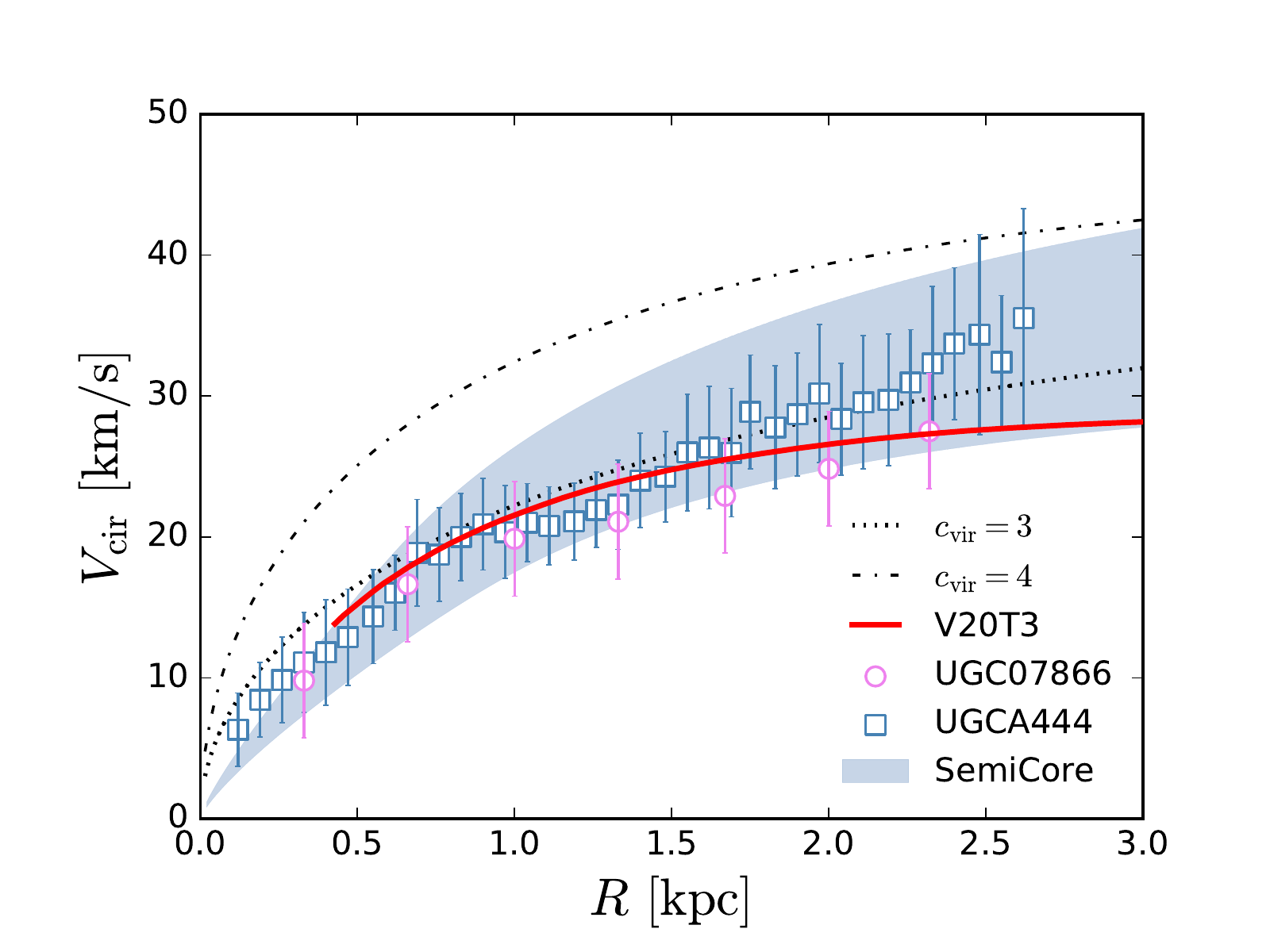}  
\includegraphics[width=0.49\linewidth]{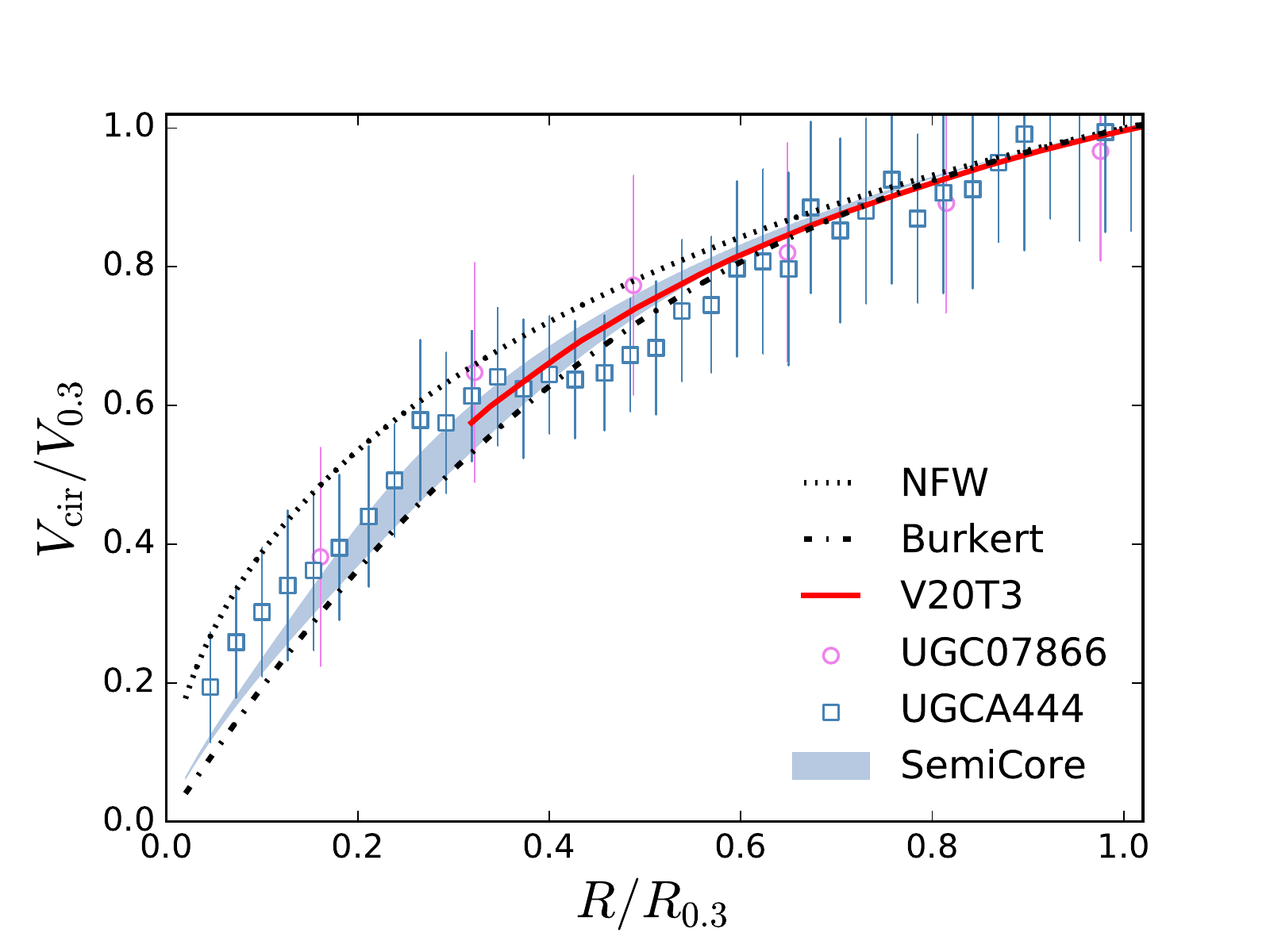}  
\caption{Comparison between DDM rotation curves with data for two dwarf galaxies from the SPARC database~\citep{lelli2016}: UGC07866 (violet circle) and UGCA444 (blue square). For the observational data, baryon contributions have been subtracted and errorbars are only shown for observational uncertainty. The red solid lines show the $\rm{V}20\rm{T}3$  curves for $r\geqslant r_{\rm{rel}}$, where statistical uncertainties are negligible. The light blue region is the prediction of the \emph{SemiCore} model with the same set of decay parameters as $\rm{V}20\rm{T}3$ but two different final virial masses: $4.12\times 10^{9}$~$h^{-1}$~$M_{\odot}$ (lower bound) and $2.0\times 10^{10}$~$h^{-1}$~$M_{\odot}$ (upper bound). Left panel: comparison of absolute rotation curves. Two NFW curves with $c_{\rm{vir}} = 3$ (black dotted line) and $4$ (black dot-dashed line) are also shown for reference. They have the same virial mass as $\rm{V}20\rm{T}3$. Right panel: comparison of scaled rotation curves. The theoretical curves based on NFW and Burkert profiles are also plotted as black dotted line and dot-dashed line, respectively. For DDM curves, the systematic uncertainty for $V_{\rm{cir}}/V_{0.3}$ at $r_{\rm{rel}}$ is estimated to be about $3\%$. At larger radii, the systematic uncertainties become unimportant. We also have $2\%$ uncertainty in the value of $R_{0.3}$, the effect of which on $V_{\rm{cir}}/V_{0.3}$ can also be safely neglected.}
\label{fig:sRC_obs}
\end{figure}

The flattening of the central density profile of dwarf halos is needed to resolve the \emph{core-cusp} problem of CDM. We show DDM and CDM's rotation curves together with observational data in Figure~\ref{fig:sRC_obs}. Halo $\rm{V}20\rm{T}3$ is a representative for the flattened DDM rotation curves. Two dwarf galaxies UGC07866 and UGCA444 are selected from the Spitzer Photometry $\&$ Accurate Rotation Curves (SPARC) database~\citep{lelli2016} for comparison as they have roughly the same mass as the halo $\rm{V}20\rm{T}3$~\citep{li2020}. In order to cover the possible mass range where the two SPARC galaxies reside, we use \emph{SemiCore} to calculate the rotation curves of two DDM halos, with current virial masses of $4.12\times 10^{9}$~$h^{-1}$~$M_{\odot}$ and $2.0\times 10^{10}$~$h^{-1}$~$M_{\odot}$, respectively. The two \emph{SemiCore} runs both use $V_{k}=20.0$~km~s$^{-1}$ and $\tau^{\ast} = 3.0$ $\mathrm{Gyr}$. Their initial halos are set to follow the CDM $c-M$ relation~\citep{diemer2018}. For the CDM rotation curve, we assume an NFW profile and a fixed virial mass $4.22\times 10^{9}$~$h^{-1}$~$M_{\odot}$, the same as the halo $\rm{V}20\rm{T}3$. In the left panel of Figure~\ref{fig:sRC_obs}, the DM contributed rotation curves of dwarf galaxy UGC07866 and UGCA444 are shown with observational uncertainties. The simulated or modelled DDM and CDM's rotation curves have been scaled by a factor of $\sqrt{1-f_b}$, where $f_b$ is the cosmic baryon fraction of global matter density. It shows that the DDM rotation curves naturally follow the observational data while the CDM ones give good fit to data only with a small concentration, which is unusual for halos formed in CDM N-body simulations~\citep{dutton2014}. The data points at small radii ($R \lesssim 0.7$ $\rm{kpc}$) favour DDM curves as the NFW curves systematically overpredict the rotation velocities there. In the right panel of Figure~\ref{fig:sRC_obs}, we plot the corresponding scaled rotation curves. With considerably large uncertainties, the observational data distribute between the NFW and Burkert curves at small radii, in good agreement with the DDM scaled rotation curves for dwarf galaxies. Our results agree with~\citet{sanchez-salcedo2003} in that the \emph{core-cusp} problem can be solved in the DDM model with a recoil velocity $V_{k}$ smaller than the typical escape velocities of dwarf halos, provided that the decay half-life $\tau^{\ast}\lesssim 7.0$ $\mathrm{Gyr}$.\

\begin{figure}[t]
\centering 
\includegraphics[width=0.7\linewidth]{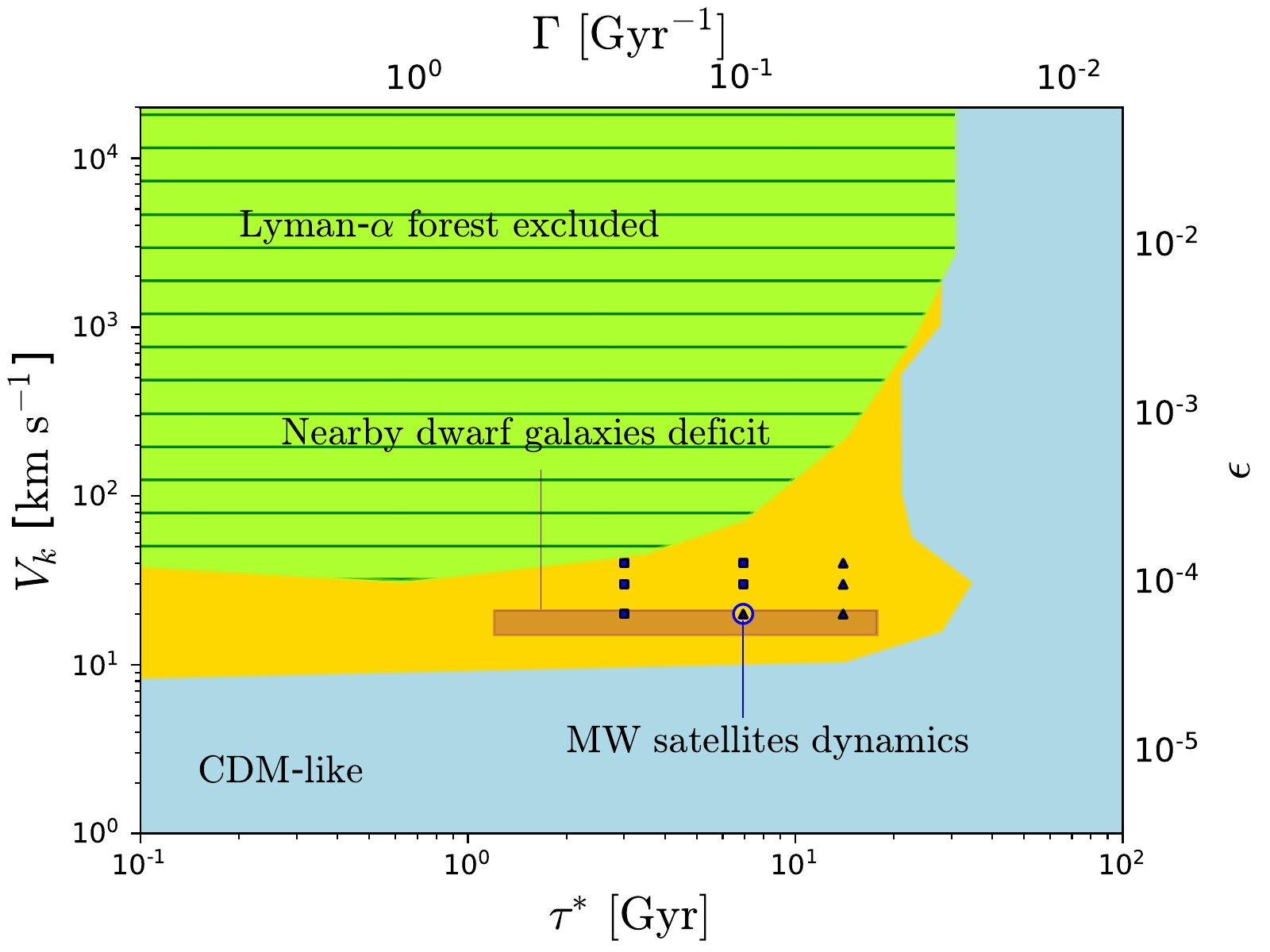}  
\caption{Constrained DDM parameter space. The green-yellow hatched region is ruled out by the Lyman-$\alpha$ forest data~\citep{meiyu2013}. The DDM model in the light-blue region behaves like the CDM model in describing the cosmic structure formation. The gold region is outlined by~\citet{peter2010c} and is interesting for small-scale problems of CDM. The blue shapes mark the $9$ level-$12$ zoom-in dwarf halos in this work, with the triangle and square indicating cuspy and cored halo density profiles, respectively. For the point $V_{k}=20.0$~km~s$^{-1}$ and $\tau^{\ast} = 6.93$ $\mathrm{Gyr}$, the dynamics of Milky-Way satellite galaxies in the DDM model is explored in~\citet{meiyu2013}. The brown strip is given by~\citet{abdelqader2008} where the deficit of dwarf galaxies in our local group can be accounted for in the DDM model.}
\label{fig:paramspace}
\end{figure}

In Figure~\ref{fig:paramspace}, we show the positions of the $9$ zoom-in dwarf halos in the DDM parameter space. They reside inside a region, shown in gold, where dark matter decays have prominent effects on the number density and inner structure of dwarf galaxies~\citep{peter2010c}. The $5$ zoom-in halos with a much more flattened central density are indicated by blue squares and the remaining $4$ by blue triangles. For $V_{k}=20.0$~km~s$^{-1}$ and $\tau^{\ast} = 6.93$ $\mathrm{Gyr}$, \citet{meiyu2014} have run a zoom-in simulation on a Milky-Way sized host halo with the PMK10 algorithm. They found that the circular velocity profiles of the $15$ most massive subhalos pass through most of the data points from the $9$ classical Milky-Way dSphs, and therefore the \emph{too-big-to-fail} problem~\citep{boylan-kolchin2011} is potentially resolved. The brown strip is the parameter region, shown by~\citet{abdelqader2008} using a semi-analytic model that incorporates dark matter decays in the hierarchical formation history of dark matter halos, that can account for the deficit of dwarf galaxies in our local group, a puzzle closely related to the \emph{missing-satellites} problem~\citep{klypin1999, moore1999b}. It can be seen that several different CDM problems can be solved by a common parameter subspace in the DDM model.\

\begin{figure}[t]
\centering
\includegraphics[width=0.7\linewidth]{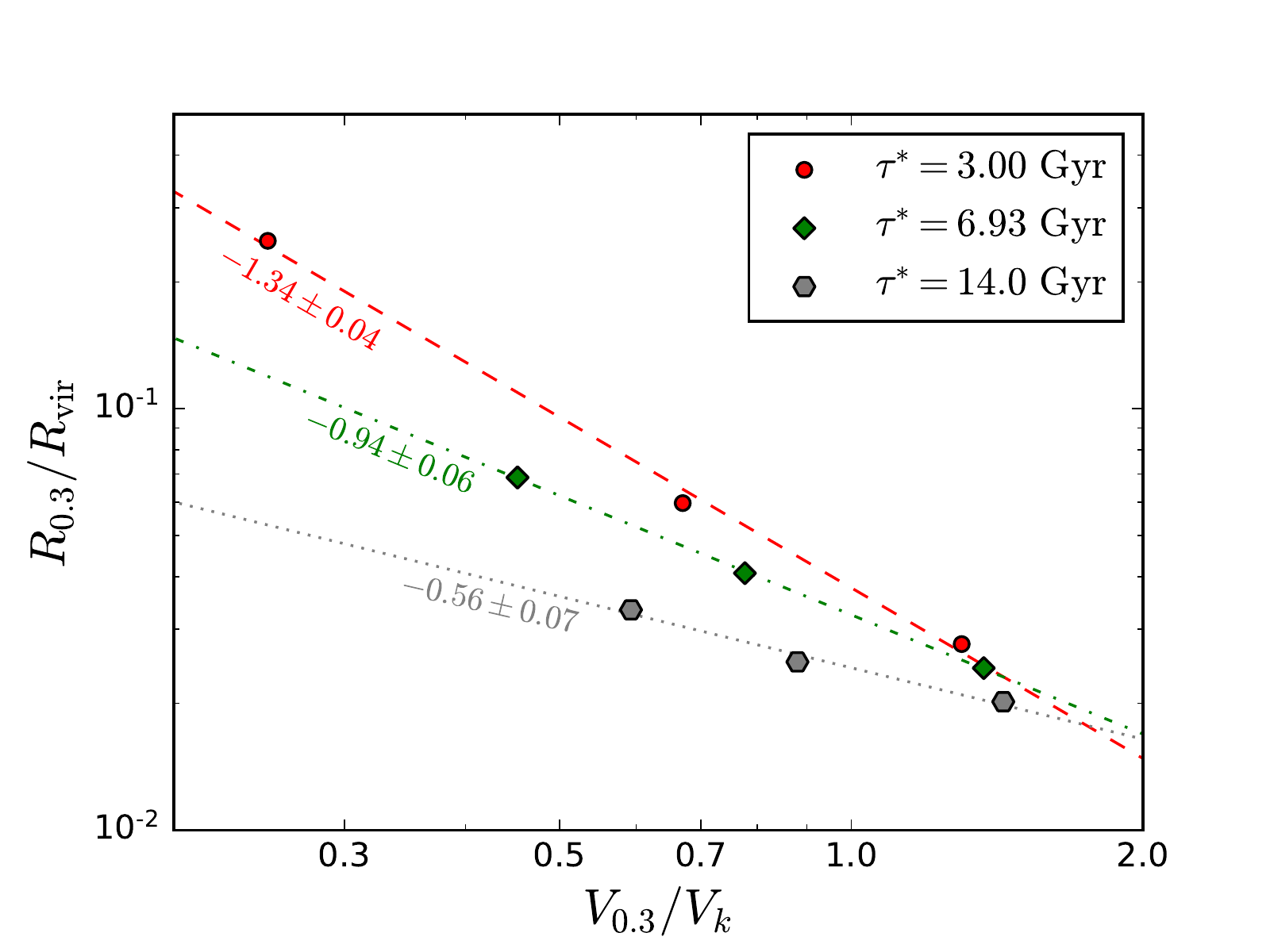}  
\caption{Relation between $R_{0.3}$ and $V_{0.3}$ for DDM zoom-in dwarf halos with the same half-life $\tau^{\ast}$. The red circles, green diamonds and gray hexagons refer to halos with $\tau^{\ast} = 3.00,$ $6.93,$ and $14.0$ $\rm{Gyr}$, respectively. There is $2\%$ uncertainty in $R_{0.3}$ and $0.6\%$ uncertainty in $V_{0.3}$. The best-fit power law curves for data points with $\tau^{\ast} = 3.00,$ $6.93,$ and $14.0$ $\rm{Gyr}$ are shown with red dashed, green dot-dashed and dotted line respectively. The best-fit power indices are also shown near each line for reference.}
\label{fig:scaling}
\end{figure}

In Figure~\ref{fig:scaling}, we show that $R_{0.3}$ and $V_{0.3}$ have a power-law relation for DDM zoom-in dwarf halos:
\begin{equation}\label{eq:scaling_relation}
                \frac{R_{0.3}}{R_{\rm{vir}}} \propto \left(\frac{V_{0.3}}{V_{k}}\right)^{-\beta(\tau^{\ast})},
\end{equation}
where the power index $\beta$ increases as $\tau^{\ast}$ decreases. Given the values of $R_{0.3}/R_{\rm{vir}}$ and $V_{0.3}$, both being observable quantities, a relation between $V_{k}$ and $\tau^{\ast}$ can be derived from relation~(\ref{eq:scaling_relation}), implying a possible degeneracy in the DDM parameter space as far as the halo's inner structure is concerned. As $V_{k}$ and $\tau^{\ast}$ are constants in the DDM model, from relation~(\ref{eq:scaling_relation}) we also find
\begin{equation}\label{eq:mass_dependence}
        \frac{V_{0.3}}{V_{\rm{vir}}} \propto \left(\frac{R_{0.3}}{R_{\rm{vir}}}\right)^{-1/\beta}M_{\rm{vir}}^{-1/3}.
\end{equation}
Since dwarf halos have a narrow virial mass spectrum and equation~(\ref{eq:mass_dependence}) depends only weakly on $M_{\rm{vir}}$, an approximate universal density profile is expected. The observed value of $\beta$ can then be used to measure $\tau^{\ast}$. We will study this possibility in a future work.

\section{Conclusions}\label{sc:sum}
In this work we improved the DDM N-body algorithm by combining the advantages of the PMK10 and CCT15 algorithms. The new algorithm outperforms PMK10 in accuracy while demands much less computing resources than CCT15. Same as CCT15, the new algorithm samples dark matter decays only at certain times and evolves the whole N-body system in a collisionless way at other times. This feature enables the algorithm to be implemented in a plugin module called DDMPLUGIN, which can be used with any CDM N-body code.\

We carried out high-resolution cosmological N-body simulations to study the density profiles of dwarf halos with the new DDM algorithm. Good numerical convergence was achieved, and we succeeded to resolve the halo structure robustly down to about $700$ pc. Compared to CDM counterparts, DDM dwarf halos have lower mass concentration and shallower density profile at the inner region. Adopting the orbit-superposition method for mass profile construction, we developed a simplified semi-analytic model for the DDM halo mass profile, which features rosette orbits for daughter particles and incorporates effects of dark matter decays and adiabatic expansion. Although simple, the model predicts DDM halo mass profiles that agree semi-quantitatively with resolved simulation profiles. It therefore illustrates clearly the physics mechanisms involved in the transformation from cusp to core density profiles.\

We also calculated the scaled rotation curves for DDM simulation halos and compared them with $2$ dwarf galaxies from the SPARC database. The shape of the DDM rotation curve is shallower than that based on the NFW profile but steeper than the Burkert's. The two SPARC dwarf galaxies favour the DDM shape of the rotation curve. Furthermore, we show that there is an approximate universal power-law relation between $V_{0.3}/V_{\rm{vir}}$ and $R_{0.3}/R_{\rm{vir}}$ for dwarf halos, which can be used to extract DDM parameters from observation data. Together with previous studies, this work supports the DDM cosmology, which keeps the success of CDM at large scale and reconciles the differences between observations and predictions from N-body simulations at small scale.

\acknowledgments
We thank Volker Springel for offering us the P-GADGET3 code. We thank the anonymous referee for the constructive comments. We also benefit from the discussions with Hantao Liu, Sze-Him Lee, Kiu-Ching Leung and Wai-Cheong Lee. We thank Hoi-Tim Cheung and Hoi-Tung Yip for helps in checking the \emph{SemiCore} results. We acknowledge the support of the CUHK Central High Performance Computing Cluster, on which the computation in this work has been performed. We have used the public python package COLOSSUS~\citep{colossus} and the NASA's Astrophysics Data System. The work described in this paper was partially supported by a grant from the Research Grants Council of the Hong Kong Special Administrative Region, China (Project No. AoE/P-404/18).

\appendix
%
\section{DDMPLUGIN implementation} \label{app:ddmplugin}
\subsection{Decay of mother particles}
At each breakpoint, the mother simulation particles will decay and release auxiliary daughter particles. Due to the random nature of dark matter decay, the N-body mother particles do not receive velocity kicks when decay occurs. Only their masses are affected and reduced by $m_{\rm{aux}}$:
\begin{equation}
                                      m'_{\rm{mom}} = m_{\rm{mom}} - m_{\rm{aux}},
\end{equation} 
where $m_{\rm{mom}}$ and $m'_{\rm{mom}}$ are the masses of a mother simulation particle just before and after the decay. Other information associated with the post-decay mother are exactly the same as those of the pre-decay mother. As for the newly generated auxiliary daughter particles, the decay kicks them away from their mother particles:
\begin{equation}
                             \boldsymbol{x}_{\rm{aux}} = \boldsymbol{x}_{\rm{mom}},
\end{equation}   
\begin{equation}
                         \boldsymbol{v}_{\rm{aux}} = \boldsymbol{v}_{\rm{mom}} + V_{k}\boldsymbol{n},
\end{equation}
where $V_{k}$ is the recoil velocity, $\boldsymbol{n}$ is a unit vector pointing to a random direction. The IDs of the auxiliary daughters are assigned as follows:
\begin{equation}
                            I_{\rm{aux}} = N_{\rm{tot},\rm{ini}} + I_{\rm{ini}} + I_{\rm{off}},
\end{equation} 
where $N_{\rm{tot},\rm{ini}}$ is the total number of particles in the initial condition, $I_{\rm{ini}}$ is the smallest particle ID taken by the simulation, and $I_{\rm{off}}$ is an integer ranging from $0$ to $N-1$. Each auxiliary daughter particle gets a different value of $I_{\rm{off}}$ randomly such that their IDs are different from each other. The masses of auxiliary daughter particles are determined by the decay half-life $\tau^{\ast}$ and the number of breakpoints $f_s$:
\begin{equation}
            m_{\rm{aux}} = m_{\rm{mom},\rm{ini}}\frac{1-\exp{\left[-\ln{(2)}t/\tau^{\ast}\right]}}{f_s},
\end{equation}
where $m_{\rm{mom},\rm{ini}}$ is the initial mass of a mother particle and $t$ is the time span of the whole simulation.

\subsection{Auxiliary-permanent transition}
The auxiliary-permanent transition of a daughter simulation particle is implemented as follows in the DDMPLUGIN module. First, the particle ID $I_{\rm{aux}}$ of an auxiliary daughter is decomposed into a pair of integers $(q,p)$:
\begin{equation}
                                        I_{\rm{aux}} = qf_s + p,
\end{equation}
where $q$ is the quotient of $I_{\rm{aux}}$ by $f_s$ and $p$ the remainder. Auxiliary daughters satisfying $p < n_f$ survive and are flagged as permanent daughters. Auxiliary daughters with $p \geqslant n_f$ are eliminated from the simulation. This scheme ensures that the survival fraction $\eta$ is $n_f/f_s$. Once an anxiliary daughter is flagged to be permanent, its mass and ID have to be modified while its position and velocity are retained. The particle mass needs modification such that the daughter particles' total mass is unaffected by the transition:
\begin{equation}
                                 m_{\rm{pmt}} = m_{\rm{aux}}\frac{f_s}{n_f},
\end{equation}
where $m_{\rm{pmt}}$ and $m_{\rm{aux}}$ are the particle masses of permanent and auxiliary daughters, respectively. A new ID $I_{\rm{pmt}}$ is assigned to the permanent daughter in order to distinguish it from its pre-transition auxiliary state:
\begin{equation} \label{eq:pmt_id_assignment}
                            I_{\rm{pmt}} = N_{\rm{tot}} + I_{\rm{ini}} + (q-q_{\rm{min}})n_f + (p-p_{\rm{min}}),
\end{equation}
where $N_{\rm{tot}}$ is the total number of particles in the simulation just before the transition, $I_{\rm{ini}}$ is the smallest particle ID taken by the simulation, $(q_{\rm{min}},p_{\rm{min}})$ is the integer pair derived from the minimum ID of the auxiliary daughters that are flagged to be permanent daughters. The ID assignment scheme~(\ref{eq:pmt_id_assignment}) ensures that the particle IDs in the simulation are continuous and there is no spatial bias in the selection process. Position and velocity are invariant under the auxiliary-permanent transition:
\begin{equation}
                                \boldsymbol{x}_{\rm{pmt}} = \boldsymbol{x}_{\rm{aux}}, 
\end{equation}
\begin{equation}
                                 \boldsymbol{v}_{\rm{pmt}} = \boldsymbol{v}_{\rm{aux}}.
\end{equation}
The auxiliary-permanent transition will also be applied when the state of the system is output at certain simulation times. This operation keeps the consistency that only permanent daughter particles appear in the output file(s) of the simulation.

\subsection{Gravitational softening length}
In our implementation of the DDMPLUGIN module, the mother simulation particles, auxiliary daughter particles and permanent daughter particles are all labelled as halo particles, which are type $1$ according to Gadget's classification. As they have the same particle type, their gravitational softening lengths are also the same.

%
\section{Numerical tests} \label{app:nut}
\subsection{Transfer function}\label{app:cts}
\begin{figure}[t]
\centering
\includegraphics[width=0.45\linewidth]{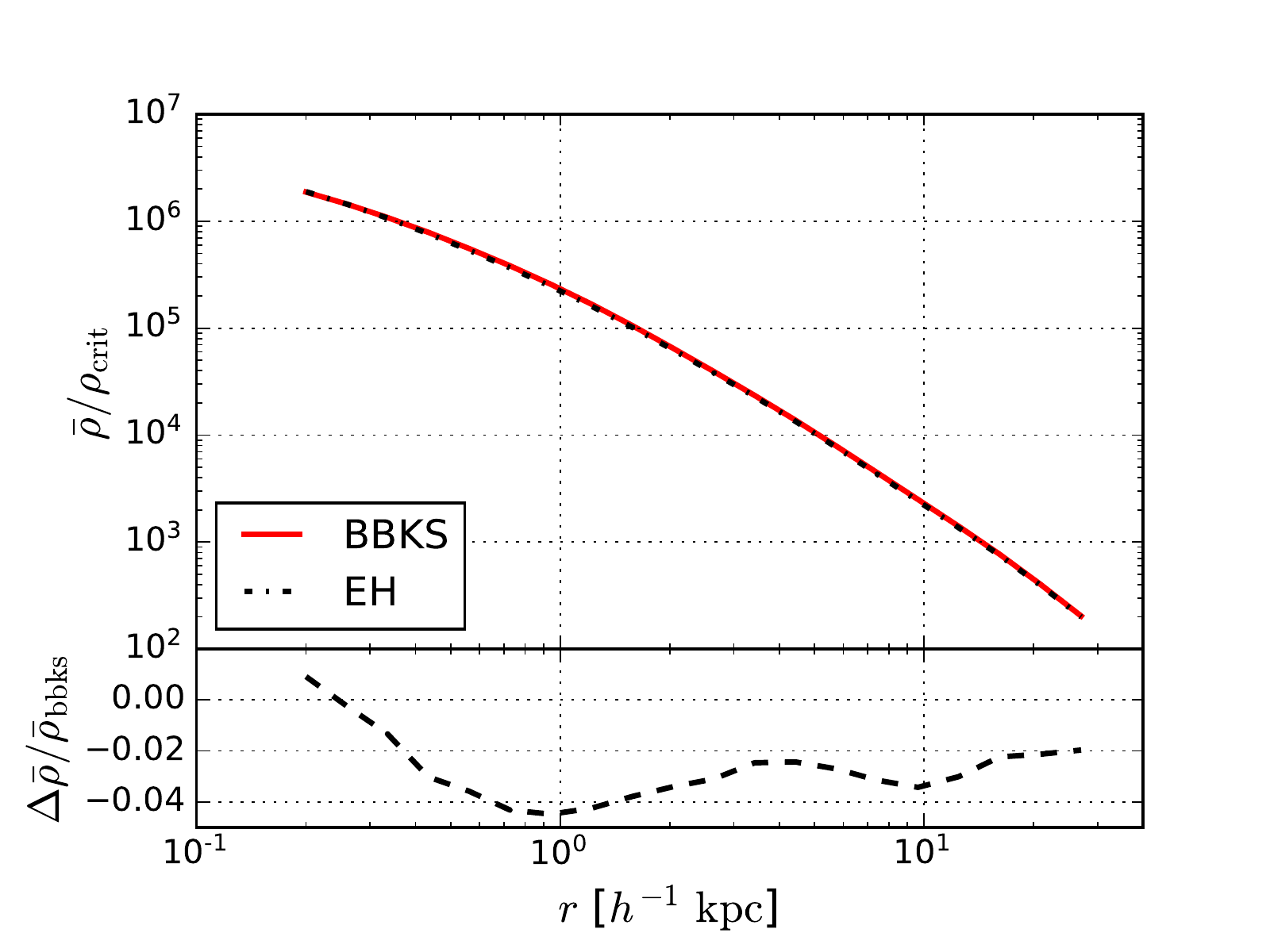}   
\includegraphics[width=0.455\linewidth]{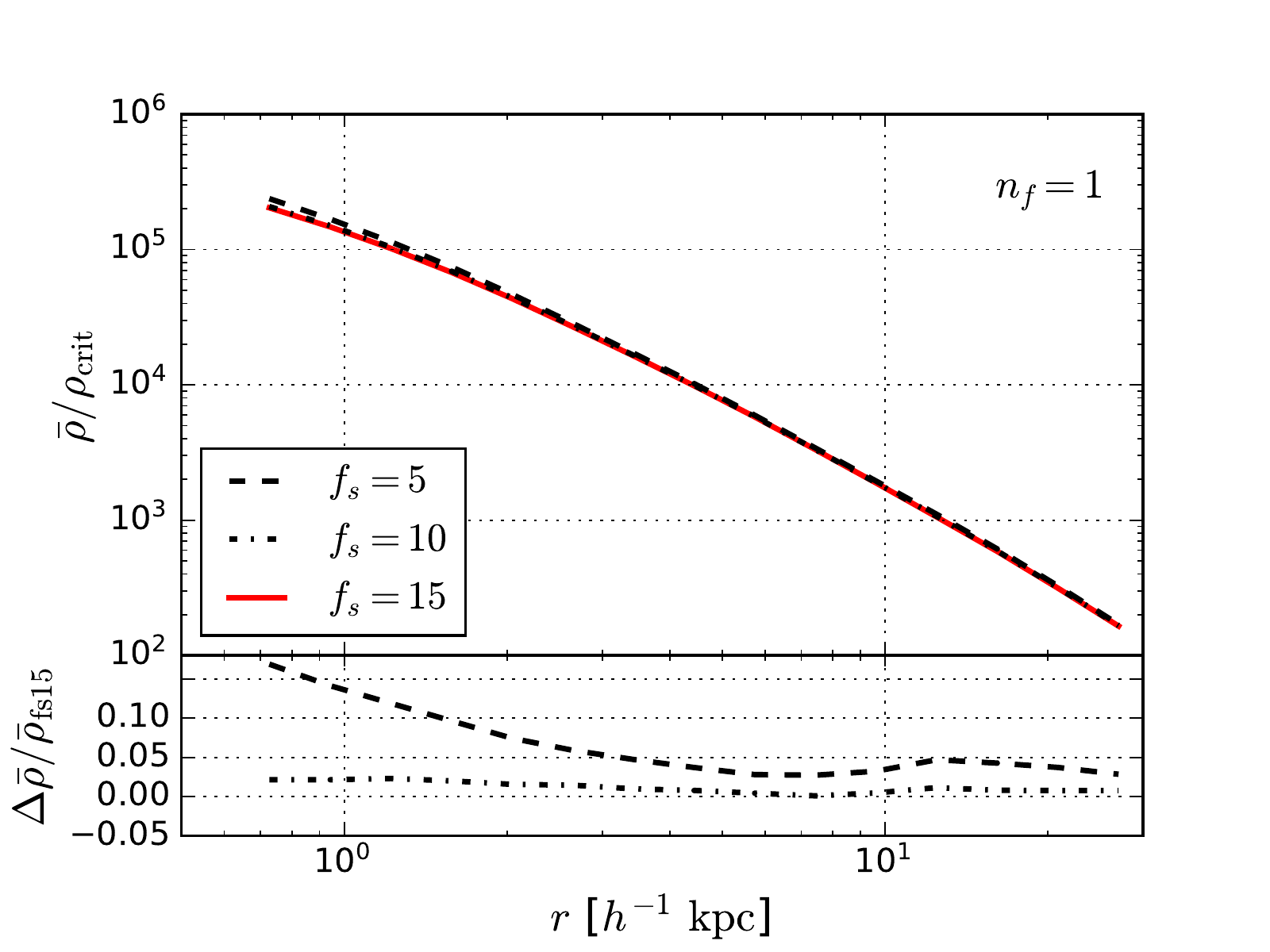}
 \caption{Left panel: Effects of transfer function variation on dark matter halo average density profile $\bar{\rho}(r)$. Densities are normalized by the current critical density $\rho_{\rm{crit}}$. The red solid line is for the BBKS transfer function while the black dot-dashed line is for the Eisenstein-Hu transfer function, and the black dashed line shows the relative differences between them normalized by BBKS's results. The average density profile is not affected by the two-body relaxation for radii larger than $0.6$~$h^{-1}$~kpc. Right panel: Effects of the numerical parameter $f_s$ on $\bar{\rho}(r)$. DDM simulations for this test all use $V_k = 20.0$~km~s$^{-1}$, $\tau^{\ast} = 3.0$ Gyr and $n_f=1$. All profiles are measured at redshift $0$ and plotted down to the inner-most resolved radii. The black dashed, dot-dashed, and red solid lines represent profiles for $f_s=5$, $10$, $15$ respectively. The profiles with $f_s=15$ serve as the baselines for comparison.}
\label{fig:nu-test}
\end{figure}

Here we test the effects of uncertainties in the transfer function on the dark matter halo's average density profile $\bar{\rho}(r)$. Two different transfer functions, the default BBKS and Eisenstein-Hu~\citep{eisenstein1998} with baryonic features, are used in the test. We run level-$11$ resolution CDM zoom-in simulations and measure the resulting profiles at $z=0$. The results are shown in the left panel of Figure~\ref{fig:nu-test}. The uncertainties on the average density profile are well within $4\%$ for most resolved radii. Hence the choice of transfer functions has little impact on our results. 

\subsection{$f_s$}\label{app:fs_test}
We used three values to test the effects of $f_s$ on the halo average density profile $\bar{\rho}(r)$: $f_s = 5, 10$ and $15$ with $n_f$ being $1$. The decayed mass fraction per phase are $19.2\%$, $9.59\%$ and $6.39\%$ for $f_s=5, 10$ and $15$, respectively. From the right panel of Figure~\ref{fig:nu-test}, it can be seen that the average densities near a halo's center are more easily affected by varying $f_s$. It is clear that convergence can be achieved by increasing the value of $f_s$. The relative differences in the average density profiles between $f_s=10$ and $f_s=15$ are within $3\%$.

\bibliography{refs}
\bibliographystyle{aasjournal}



\end{document}